\documentclass[a4paper,12pt]{article}
\pdfoutput=1 
\usepackage{graphicx}% Include figure files
\usepackage{bm}% bold math
\def\be{\begin{equation}}
\def\ee{\end{equation}}
\def\bea{\begin{eqnarray}}
\def\eea{\end{eqnarray}}
\usepackage{latexsym}
\usepackage{epsfig,amssymb,euscript,mathrsfs}
\usepackage{amsmath}
\usepackage{array,calc,epsfig}

\usepackage[noadjust]{cite}

\usepackage{chngcntr}

\usepackage[pdfpagelabels]{hyperref}

\usepackage{color}

\usepackage{color}
\newcommand{\Tr}{{\rm Tr}}

\begin{document}
\pagestyle{plain}
\baselineskip=15.5pt
\setcounter{page}{1}
\newfont{\namefont}{cmr10}
\newfont{\addfont}{cmti7 scaled 1440}
\newfont{\boldmathfont}{cmbx10}
\newfont{\headfontb}{cmbx10 scaled 1728}
\renewcommand{\theequation}{{\rm\thesection.\arabic{equation}}}
\renewcommand{\thefootnote}{\arabic{footnote}}
\vspace{1cm}
\begin{titlepage}
\vskip 2cm
\begin{center}
{\Large{\bf Hall Droplet Sheets in Holographic QCD}}
\end{center}

\vskip 10pt
\begin{center}
Francesco Bigazzi$^{a}$, Aldo L. Cotrone$^{a,b}$, Andrea Olzi$^{b}$
\end{center}
\vskip 10pt
\begin{center}
\vspace{0.2cm}
\textit{$^a$ INFN, Sezione di Firenze; Via G. Sansone 1; I-50019 Sesto Fiorentino (Firenze), Italy.}\\
\textit{$^b$ Dipartimento di Fisica e Astronomia, Universit\'a di Firenze; Via G. Sansone 1;\\ I-50019 Sesto Fiorentino (Firenze), Italy.}
\vskip 20pt
{\small{bigazzi@fi.infn.it, cotrone@fi.infn.it, andrea.olzi@stud.unifi.it} }

\end{center}

\vspace{25pt}

\begin{center}
 \textbf{Abstract}
\end{center}

\noindent 

In single-flavor QCD, the low energy description of baryons as Skyrmions is not available.
In this case, it has been proposed by Komargodski that  baryons can be viewed as kinds of charged quantum Hall droplets, or ``sheets''.
In this paper we propose a string theory description of the sheets in single-flavor holographic QCD, focusing on the Witten-Sakai-Sugimoto model.
The sheets have a ``hard'' gluonic core, described by D6-branes, and a ``soft'' mesonic shell, dual to non-trivial D8-brane gauge field configurations.
We first provide the description of an infinitely extended sheet with massless or moderately massive quarks.
Then, we construct a semi-infinite sheet ending on a one-dimensional boundary, a ``vortex string''.
The holographic description allows for the precise calculation of sheet observables.
In particular, we compute the tension and thickness of the sheet and the vortex string, and provide their four dimensional effective actions.

\end{titlepage}

\tableofcontents

\section{Introduction}\label{intro}
\setcounter{equation}{0}
In the effective field theory of QCD, baryons can be described as solitons of the chiral Lagrangian equipped with the Skyrme term. This picture is well justified in the large $N$ limit, where the baryon mass, linear in $N$, scales like the inverse squared meson coupling. However, this description does not hold if the number of flavors reduces to $N_f = 1$. In this case, in fact, the group manifold of the chiral Lagrangian trivializes and it is not possible to write down a conserved baryonic charge. Hence, one-flavored baryons, which have spin $N/2$, require an alternative low-energy description. 

An intriguing proposal has been put forward by Komargodski in \cite{Komargodski:2018odf}, building on earlier works such as \cite{Kogan:1993yw,Gabadadze:2000vw,Forbes:2000et,Son:2000fh,Gabadadze:2002ff}. In large $N$ QCD with just one quark whose mass is parametrically lower than the QCD scale, the would-be $\eta'$ meson is a light particle. The $\eta'$ is a periodic field being the phase of the chiral condensate, and its effective theory admits extended $\eta'$ configurations, $(2 + 1)$-dimensional sheets, that interpolate from the $\eta' =0$ vacuum on one side to the (equivalent) $\eta' =2\pi$ vacuum on the other side. Around these sheets the $\eta'$ has a $2\pi$ monodromy. The sheets can end on a boundary. The related $\eta'$ configuration cannot be smooth everywhere, because the monodromy is violated outside the boundary. Thus, the $\eta'$ field should be singular and the minimal singularity that must exist is of the form of the boundary. As a consequence, a locus of discontinuity for the $\eta'$ field arises, which can be thought of as a Dirac sheet, where it jumps from $0$ to $2\pi$. In \cite{Komargodski:2018odf} it has been suggested that one-flavored baryons can be classically thought of as pancake-shaped sheets, with a $N$-independent size and a mass proportional to $N$.

Another key element of this construction is that the effective field theory on the sheet is a $U(1)_N$ topological field theory which describes the bulk physics of a fractional quantum Hall state. Then, the sheets can be interpreted as baryons by coupling a background baryonic gauge field to the theory on the sheet. Indeed, this coupling is precisely like the coupling of the electromagnetic gauge field to the emergent gauge field in the fractional quantum Hall effect (FQHE). If the sheet has a boundary, due to the bulk-boundary correspondence of FQHE, there can be edge excitations which carry a conserved baryon number.
The charge of these modes forbids the sheet to collapse, stabilizing its size.
Other studies of this proposal can be found e.g.~in \cite{Ma:2019xtx,Karasik:2020pwu,Ma:2020nih,Karasik:2020zyo,Kitano:2020evx,Nastase:2022pts}.

The low-energy $\eta'$ theory is incomplete: there is a cusp singularity in the potential which prevents us to perform explicit calculations in this framework. The physical interpretation of the cusp is that in correspondence of the sheet some heavy fields rearrange. These heavy fields are likely to be glueballs, which are not taken into account in the chiral Lagrangian.  

This paper aims at exploring and testing the above proposal in top-down holographic QCD, namely within the Witten-Sakai-Sugimoto (WSS) model \cite{Witten:1998zw,Sakai:2004cn}. This is a non-supersymmetric $SU(N)$ gauge theory in $3+1$ dimensions, coupled to $N_f$ fundamental flavors and a tower of massive (Kaluza-Klein) adjoint fields. The Yang-Mills plus adjoint field sector is realized as the low energy theory of a stack of $N$ D4-branes wrapped on a circle. At strong coupling and in the planar limit, it is holographically described by a dual Type IIA supergravity theory on a specific background, which remarkably accounts for confinement and mass gap formation in the dual field theory. Fields in the fundamental representation, i.e. quarks, are obtained by introducing $N_f$ pairs of D8-$\overline{\rm D}$8-branes, which, in the $N_f\ll N$ limit, can be treated as probes on the above mentioned background. To minimize their energy these branes form a unique U-shaped configuration, geometrically realizing the spontaneous breaking of chiral symmetry. The effective action on the flavor branes describes a $U(N_f )$ Yang-Mills-Chern-Simons theory in five curved dimensions. Mesons arise as excitations of the $U(N_f)$ gauge field, i.e.~of open strings on the flavor branes, while baryons (wrapped D4-branes in the zero-size limit, following \cite{wittenbaryon}) are identified with instantonic configurations of the $U (N_f)$ field \cite{SS-barioni}. Remarkably, the same problem of finding a baryon charge in the low energy description of $N_f=1$ large $N$ QCD arises also in the holographic model \cite{Seki:2008mu}. This makes even more urgent to find a way to construct an alternative low energy description of the corresponding baryons.

In this paper we propose a holographic realization of the sheet in the one-flavored WSS model. This is achieved by adding an extra D6 probe brane, wrapped on the four-sphere of the gravitational background, and extended in $(2+1)$ Minkowski directions. The world-volume of the D6-brane, after a dimensional reduction on the four-sphere, precisely supports the expected $U(1)_N$ Chern-Simons theory realizing the bulk physics of the FQHE. In this picture, the sheet can be viewed as a wall with a hard (gluonic) D6-brane core, corresponding to the heavy fields associated to the singularity in the chiral Lagrangian, and a soft shell (a mesonic cloud) coming from the non-trivial $\eta'$ profile. Indeed, the D6-brane can be viewed as a source for the Maxwell gauge field living on the D8-brane, whose excitations are identified with the $\eta'$ meson.

The holographic construction makes it transparent that the dynamical gauge field on the sheet is not of mesonic nature, rather it sources some mesonic profiles through the boundary.
Some discussions on this topic appear in \cite{Karasik:2020pwu,Karasik:2020zyo,Kitano:2020evx}.
Nevertheless, as we will briefly discuss in section \ref{secLesson}, it is possible that in QCD with light quarks the dynamical field is Higgsed.
In this case, one should be able to describe the sheet entirely in terms of mesonic degrees of freedom.
We will not explore this possibility in this paper, leaving this interesting issue for a future study. 

To simplify our analysis, we will focus on two  configurations: the infinite and the semi-infinite sheet with a straight boundary. In the first case, as expected, we find a regular profile for the $\eta'$ field both in the massless and the massive quark case. We compute the tension and the thickness of the sheet considering the contributions from the hard core and the mesonic cloud. In the second case, the straight boundary can be viewed as a linear distribution of magnetic charge in the five-dimensional Maxwell-Chern-Simons effective theory on a curved background. The corresponding gauge field can be found semi-analytically, through a series expansion, and numerically. From this solution we extract the effective action for the transverse fluctuations of the boundary, and the tension, which scales parametrically as expected in \cite{Komargodski:2018odf}. Then, we can compute the $\eta'$ profile which now has two different features depending on the domain. In fact, it is continuous (in analogy with the infinite sheet case) through the D6-brane, while it is discontinuous outside the sheet, enforcing the monodromy of the $\eta'$ meson around the wall. 
Finally, we will show that it is possible to turn on, with no energy cost, a chiral component of the gauge field on the boundary of the sheet; this is possibly connected to the chiral edge mode  on the boundary of the Hall droplet.

The paper is organized as follows.
After a brief review of the Witten-Sakai-Sugimoto model in section \ref{secbackground}, we describe our proposal for the dual picture of the sheet in section \ref{secSheetinWSS}.
Then, we move on to the construction of the infinitely extended sheet, both with massless and massive quarks, in section \ref{sec:infinite}.
Section \ref{secSemiInfinite} contains the main results of this paper, concerning a semi-infinite sheet terminating on a string-like boundary.
After a warm-up in flat space in section \ref{secFlat}, we provide the complete solution and its properties in section \ref{sec:semicurved}.
We conclude with a summary of our results and some directions for future investigations in section \ref{secConclusions}.

\section{The background}\label{secbackground}
\setcounter{equation}{0}
The WSS model of holographic QCD is based on the type IIA background \cite{Witten:1998zw}
\bea \label{background}
ds^2&=&\left(\frac{u}{R}\right)^{3/2} \left(dx^\mu dx_\mu + f(u)d x_4^2 \right)+ \left(\frac{R}{u}\right)^{3/2}\frac{du^2}{f(u)}
+R^{3/2}u^{1/2}d\Omega_4^2\ ,\nonumber \\     f(u)&=&1-\frac{u_0^3}{u^3}\ ,  \qquad     e^\Phi=g_s\frac{u^{3/4}}{R^{3/4}}\ , \qquad F_4 = 3 R^3 \omega_4 \,,  
\eea
where $\mu=(0,1,2,3)$, the radial variable $u$ has dimensions of length and ranges in $[u_0,\infty)$, the circle $S_{x_4}$ has length $\beta_4$, $R=(\pi g_s N)^{1/3}l_s$, $\omega_4$ is the volume form of $S^4$ (which has volume $V(S^4)=8\pi^2/3$) and absence of conical singularities gives the relation $9 u_0 \beta_4^2 = 16 \pi^2 R^3$. $\Phi$ is the dilaton and $F_4$ the Ramond-Ramond four-form field strength. Moreover $\beta_4=2\pi/M_{KK}$, where $M_{KK}$ is the Kaluza-Klein (KK) and glueball mass scale,  proportional to the dynamical scale $\Lambda$
\begin{equation}
M_{KK} \quad \leftrightarrow \quad \Lambda\,.
\end{equation}
The above background holographically accounts for confinement and mass gap formation in the dual $SU(N)$ theory. In the holographic regime, the confining string tension of the theory scales like $\lambda M_{KK}^2$, where $\lambda\equiv g_{YM}^2N = 2\pi g_s N l_s M_{KK}\gg1$. The background can be further enriched by a Ramond-Ramond one form potential $C_1$ in case one is interested in turning on a Yang-Mills $\theta$ angle \cite{Witten:1998uka}. In the following we will work at $\theta=0$. 

Flavor degrees of freedom are introduced by means of  $N_f$ D8-$\overline{\rm D}8$ branes at fixed points on the $S_{x_4}$ circle \cite{Sakai:2004cn}. These branes are usually treated in the probe approximation which is reliable if $N_f\ll N$. To minimize their energy they actually form a unique U-shaped configuration on the ($x_4, u$) cigar-like subspace, dynamically realizing the spontaneous breaking of chiral symmetry. If the D8-$\overline{\rm D}8$ branes are at antipodal (resp.~non antipodal) points on the $S_{x_4}$ circle, the tip of the U-shaped configuration is at $u=u_0$ (resp.~at $u_1>u_0$). 

The D8-brane embedding can be described in terms of the coordinates
\be
y = \frac{u^{3/2}}{\sqrt{u_0}}\sqrt{f(u)} \cos (M_{KK} x_4)\,,\qquad z = \frac{u^{3/2}}{\sqrt{u_0}}\sqrt{f(u)} \sin (M_{KK} x_4)\,.
\ee
In these coordinates the antipodal D8-brane is localized at $y=0$ and extends along $z\in (-\infty,\infty)$ with the tip at $z=0$. In units $M_{KK}=u_0=1$, $R^3 =9/4$ and the cigar metric in the new coordinates reads
\begin{equation}\label{cignewcoord}
d s^2_{(y,z)} =\frac{2}{3u^{3/2}}\left[\left(1-q\,z^2\right)d z^2 + \left(1-q\,y^2\right)d y^2 - 2zy \, q\,d z d y\right]\,,
\end{equation}
with $u=u(y,z)=(1+y^2+z^2)^{1/3}$, $q=q(y,z)=\frac{1}{y^2+z^2}\left(1-\frac{1}{u}\right)$. 

The effective action for $N_f$ D8-branes at $y=0$, after reduction on $S^4$ and to second order in derivatives reads
\begin{equation}\label{SYM}
S_f = -\kappa\int d^4x d z\, \left(\frac{1}{2}h(z)  \, \Tr\mathcal{F}_{\mu\nu}\mathcal{F}^{\mu\nu} +   k(z)  \Tr\mathcal{F}_{\mu z}\mathcal{F}^\mu_{\;\; z}\right)
 + \frac{N}{24\pi^2}\int  \omega_5(\mathcal{A}) \,,
\end{equation}
where four-dimensional indices $\mu,\nu$ are raised by the flat Minkowski metric and (in units $M_{KK}=u_0=1$)
\be
\kappa = \frac{N\lambda}{216\pi^3}\,,\quad h(z) = (1+z^2)^{-1/3}\,,\quad k(z) = (1+z^2)\,,
\ee
and 
\begin{equation}
\omega_5(\mathcal{A}) =  \Tr\left(\mathcal{A}\wedge \mathcal{F}^{ 2}- \frac{i}{2}\mathcal{A}^{ 3} \wedge\mathcal{F} - \frac{1}{10}\mathcal{A}^{ 5}\right)\;,\quad d \omega_5(\mathcal{A}) = \Tr\mathcal{F}^{  3}\,.
\label{omega5}
\end{equation}
The gauge field $\mathcal{A}\in U(N_f)$ can be decomposed in the Abelian and non-Abelian parts
\begin{equation}
\mathcal{A} = A \frac{\mathbf{1}}{\sqrt{2N_f}} + \hat A^a T^a\,,
\label{separ}
\end{equation}
where $T^a$ are the $SU(N_f)$ generators. 

Using this notation the action can be rewritten as
\begin{equation}
\begin{aligned}
S_{f} &= -\kappa\int d^4x d z\, \left(\frac{1}{2}h(z)  \, \Tr \hat F_{\mu\nu}\hat F^{\mu\nu} +   k(z)  \Tr \hat F_{\mu z}\hat F^\mu_{\;\; z}\right)
 +\\&\quad  
 -\frac{\kappa}{2}\int d^4x d z\, \left(\frac{1}{2}h(z)  \, {F}_{\mu\nu}{F}^{\mu\nu} +   k(z) {F}_{\mu z}{F}^\mu_{\;\; z}\right)
 +\\&\quad  
+\frac{N}{24\pi^2}\int  \biggl[\omega_5^{SU(N_f)}(\hat A)+ \frac{3}{\sqrt{2N_f}}{A}\,\Tr \hat F^2 + \frac{1}{2\sqrt{2N_f}}{A}\,{F}^2\biggr]\,.
\label{clac}
\end{aligned}
\end{equation}
Here $\omega_5^{SU(N_f)}$ is defined as in (\ref{omega5}), written in terms of just the non-Abelian components. It is worth noticing that it is identically zero for $N_f\le2$. 

In fact, in the following we will focus on the $N_f=1$ case, where the effective action above describes a five-dimensional Maxwell-Chern-Simons theory 
\begin{equation}
S^{N_f=1}_{f} = -\frac{\kappa}{4}\int d^4x d z\, \sqrt{-\det g}\, g^{MN}g^{RS} F_{MR}F_{NS}+\frac{N}{24\pi^2}\int \frac{1}{2\sqrt{2}}A\wedge F\wedge F
\label{clacNf1}
\end{equation}
on the curved background
\begin{equation}\label{curvedspace}
ds^2 = g(z) dx^{\mu}dx_{\mu} +  \frac{dz^2}{g(z)}\,, \qquad g(z)=(1+z^2)^{2/3}\,.
\end{equation}
It turns out that this provides an effective description for the $\eta'\sim\int{A}_z dz$ meson. 

In fact, mesons arise as gauge field fluctuations in the above effective action, while baryons (wrapped D4-branes in the zero size limit, following \cite{wittenbaryon}) are realized as instanton solutions \cite{SS-barioni} provided $N_f\ge2$. In the $N_f=1$ case we thus need an alternative description. This will be the focus of the following sections.

%%%%%%%%%%%%%%%%%%%%%%%%%%%%%%%%%%%%%%%%%%%%%%%
\section{The sheet in the WSS model}
\label{secSheetinWSS}
\begin{figure}
\center
\includegraphics[scale=0.9]{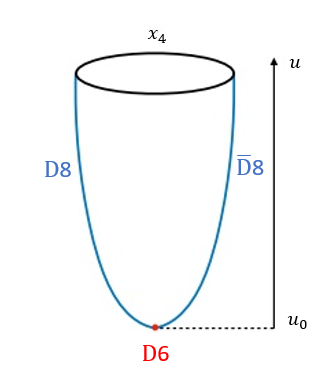}
\caption{The displacement of the D-branes on the cigar of the geometry (\ref{background}) for antipodally embedded D8-branes.}
\label{figcigar}
\end{figure}

\setcounter{equation}{0}
We propose that the holographic picture for the $N_f=1$ baryons considered in \cite{Komargodski:2018odf} contains a D6-brane wrapped on $S^4$. When the D8-branes are antipodally embedded, the D6-brane is localized at $z=0$, $y=0$ (i.e.~$u=u_0$) and can be taken at, say,  $x_3=0$ in the geometry (\ref{background}), (\ref{cignewcoord}); see figure \ref{figcigar} for the displacement of the branes on the cigar of the geometry. 
The low energy theory on the D6-brane world-volume is given by a $U(1)_N$ CS theory (plus the center of mass degrees of freedom) just because it is wrapped on the $S^4$ \cite{Argurio:2018uup}, which supports $N$ units of flux of $F_4$. The DBI action of the D6-brane gives, upon integration on the four-sphere, the three-dimensional kinetic term for the Abelian gauge field $a$ which lives on the D6-brane world-volume. Moreover, from the CS term of the D6-brane we get
\be
\frac{1}{8\pi^2}\int C_3\wedge da \wedge da = \frac{1}{8\pi^2}\int F_4 \wedge a \wedge da = \frac{N}{4\pi}\int a \wedge da\,,
\ee
where we have used $\int_{S^4} F_4 = 2\pi N$. Remarkably the above theory precisely agrees with Komargodski's proposal for the theory on the bulk of the sheet. 

This picture resembles that of D6-branes as domain walls (DW) separating different degenerate vacua in the WSS model at $\theta=\pi$ \cite{Witten:1998uka,Argurio:2018uup}.\footnote{The world-volume theory in the case of QCD has been studied in field theory in \cite{Gaiotto:2017tne}.} 
Across the domain wall there is a jump of the value of the gluon condensate while in our case there is a monodromy of $\eta'$. Notice moreover that the sheet now does not separate two different vacua since the $\eta'=0$ and the $\eta'=2\pi$ one are the same. The story goes in close analogy with the axion monodromy studied in \cite{Dubovsky:2011tu} where interesting comments on the metastability of the D6-brane configuration with a boundary on the D8-brane, can be found. 
In \cite{Dubovsky:2011tu} in fact it is argued that the D6 sheet is at best metastable, since it can decay by developing holes with an exponential suppressed rate $\Gamma \sim e^{-\lambda^2 N}$. 

As already remarked, the holographic picture of the sheet does not end with the D6-brane. The latter (which in the baryon case is expected to have a finite, disk-like shaped extension) has to be thought of as the hard core of the Hall droplet baryon. On top of this there will be a soft mesonic cloud due to the non trivial configuration of the $\eta'$ field. The latter can be explicitly found considering the whole structure of the D8-brane action in presence of a D6 source. 

The D8-brane action for $N_f=1$ admits a CS term 
\begin{equation}
\frac{1}{2\pi} \int C_7 \wedge \frac{F}{\sqrt{2}}\,,
\end{equation} 
so that a charge for a D6-brane with the same embedding as that of the source can be induced by a non trivial flux for the abelian gauge field $ F$. As we will see in the following, the D6-brane source actually induces a flux 
\be
\int F_{x_3 z} dx_3 dz = - 2\pi \sqrt{2}\,,
\label{cond1}
\ee
which provides an $\overline{\rm D}6$-brane charge precisely canceling that of the source. Hence the total D6 charge in the setup is zero. The sheet in fact can decay and thus there cannot be a net D6 charge.
 
The condition (\ref{cond1}) precisely accounts for a non-trivial profile of the $\eta'$ field. The mesonic shell of the sheet can be seen as a non-trivial solution for the Abelian gauge field on the D8-brane satisfying (\ref{cond1}). 

However, this is not enough to get the action for a point-like particle with baryonic $U(1)_B$ charge on the D8-branes. In fact defining $B_0(t) = {A}_0/\sqrt{2}$ and treating it as a time dependent perturbation, we can get such a point-particle action only if the other component ${F}_{12}$ of the field strength is turned on. Indeed from the five-dimensional CS part of the action (\ref{clacNf1}) we get
\be
\frac{N}{24\pi^2}\int \frac{1}{2\sqrt{2}}{A}\wedge\,{F}\wedge{F}=\frac{N}{24\pi^2}\int dt \,B_0(t) \int\frac{F}{\sqrt{2}}\wedge\frac{F}{\sqrt{2}}\,,
\ee
which implies that the baryon solution has to be some kind of ``Abelian instanton" with unit instanton number
\be
\int ({F}\wedge{F})_{x_1 x_2 x_3 z} dx_1 dx_2 dx_3 dz = 48\pi^2\,.
\label{conbar}
\ee
Obviously in the Maxwell-CS theory there is no solution of this kind on $\mathbb{R}^4$. The only possibility is that there is some kind of non trivial extended source. If the overlap between the sheet and the D8-branes in the $(x_1,x_2)$ plane is a disk of radius $L$  then the above condition has some hope to be satisfied. Analogously to what happens for the standard baryons, due to the five-dimensional CS coupling the above condition induces an electric field. We expect that the related Coulomb interaction is what stabilizes the sheet to some finite size $L_{stable}$.\footnote{We do not have any argument concerning the scaling of $L_{stable}$ with $\lambda$: it might be that, as for the standard baryon in the WSS model, the size is parametrically small, $L_{stable} \sim 1/(\lambda \Lambda)$, regardless of the large spin. This issue can be settled only studying the full solution.} 
In this paper we will not consider this setup, and thus we will not impose the condition (\ref{conbar}), leaving the analysis of the actual quantum Hall droplet baryon for the future. 

%%%%%%%%%%%%%%%%%%%%%%%%%%%%%%%%%%%
\subsection{Comparison with QCD}
\label{secLesson}

In the WSS model, departure from the antipodal D8-brane configuration corresponds to increasing quartic NJL-type couplings between the quarks \cite{Antonyan:2006vw}.
If this departure is large, i.e.~the configuration is far from the antipodal one, the effect of large NJL-type couplings is that the critical temperature for chiral symmetry breaking is different (larger) than that for confinement \cite{Aharony:2006da}.
Thus, this scenario is different from what is expected to happen in QCD.
Instead, if the departure from the antipodal configuration is not large, the chiral symmetry breaking and confinement temperatures are still coincident and there is no clear contradiction with QCD expectations. 

Thus, the configurations closer to QCD are the antipodal one and the moderately non-antipodal ones.
In the antipodal case, the holographic description of the sheet is a D6-brane lying on top of the D8-branes at the tip of the cigar (as in figure \ref{figcigar}).
In flat space, such a configuration is unstable towards the formation of a D8-D6 bound state, where the D6 is melt in the D8 world-volume.
This happens because the spectrum of D8-D6 strings contains a scalar.
The scalar is tachyonic if the branes are on top of each other (and there are no fluxes turned on).
The melting of the D6 corresponds to the condensation of this scalar.\footnote{It is not immediately evident how to describe the three-dimensional scalar field on the world-volume of the sheet in terms of the elementary fields of the four-dimensional theory.}
This corresponds to a complete Higgsing of the $U(1)$ gauge group of the three-dimensional theory living on the world-volume of the sheet (the D6-brane). In this scenario it should be
again possible to describe finite energy baryons. A very similar situation is found in the case of the domain-wall at $\theta=\pi$ \cite{Argurio:2018uup}.

A small departure from the antipodal configuration is sufficient to give a positive mass to the scalar, avoiding the Higgsing mechanism and so the melting of the D6-brane. Thus, there is a family of moderately non-antipodal configurations whose phase diagram is consistent with the QCD one and for which the sheet is realized with an actual D6-brane which is not melt.\footnote{We thank Ofer Aharony for exchange of ideas on this issue.} The D6 melting could be avoided also by introducing sufficiently large masses for the quarks \cite{Gabadadze:2000vw,Argurio:2018uup}.  

The condensed scalar scenario is not the one explored in \cite{Komargodski:2018odf}.
While it would be of obvious interest to explore the melted-D6 configuration, in this paper we aim at studying the original proposal in \cite{Komargodski:2018odf},  corresponding to an actual D6-brane.
In order to simplify the computations we will consider the antipodal configuration without large quark masses, aiming at revealing some basic features of the sheet.
This is obviously an oversimplification, which eventually must be lifted in future studies.

%%%%%%%%%%%%%%%%%%%%%%%%%%%%%%%%%
\section{Infinite sheet}
\label{sec:infinite}
\setcounter{equation}{0}
In this section we study the infinitely extended wall corresponding to a sheet without boundaries.
The sheet has a ``hard'', gluonic core corresponding to the D6-brane, and a ``soft'' shell from the $\eta'$ profile, i.e.~a meson cloud, as advocated e.g.~in \cite{Gabadadze:2000vw}.

%From now on we will work with $N_f=1$.
%, the extension to more flavors and/or to more D6-brane sources being straightforward.
%Also, we are going to drop the hat on the Abelian part of the gauge field for notational convenience.

\subsection{The core}

As we have already pointed out, the ``hard'', gluonic core of the sheet is holographically given by the wrapped D6-brane.
Assuming the sheet to be extended along the $x_1, x_2$ directions, its tension $T_{hard}$ can be extracted from the DBI term of the wrapped D6-brane action
\begin{equation}
S^{DBI}_{D6} = -T_6 \int d^7x e^{-\Phi} \sqrt{-g_7} \equiv -T_{hard} \int dt dx_1dx_2\,,
\end{equation}
where $T_6 = (2\pi)^{-6}\, l_s^{-7}$ and
\begin{equation}
T_{hard} = \frac{1}{\pi^3 3^6} \lambda^2 N M_{KK}^3\,.
\end{equation}
As expected, the tension scales as $N$, but is enhanced by the $\lambda^2$ factor, as usual in holographic theories.

Considering the thickness $\delta$ of a D-brane sitting at the tip of the background cigar to be given by the effective local string length $l_{s, eff}$, one gets for the thickness of the hard core of the wall (up to a numerical factor)
\begin{equation}\label{thickhard}
\delta_{hard} \sim l_{s,eff} \sim \frac{1}{\lambda^{1/2}M_{KK}}\,,
% l_s \left( \frac{R}{u_0} \right)^{3/4}
\end{equation}
which comes from the scaling $T_s\sim l_{s, eff}^{-2}\sim \lambda M_{KK}^2$ of the Yang-Mills string tension in the model. 

\subsection{The shell}
\label{secinf}

A localized D6-brane is a source for the RR $C_7$ field. 
This in turn will generate a non-trivial profile for the D8-brane mode corresponding to the $\eta'$.
The profile of the  $\eta'$, the meson cloud, constitutes the ``soft shell'' of the sheet.  

In order to study the D6-D8 setup we have in mind we should better consider the total action $S_f^{N_f=1} + S_{C_7}$ in the presence of D6-brane sources where $S_f^{N_f=1}$ is given in (\ref{clacNf1}) and 
\be
S_{C_7} = -\frac{1}{4\pi} (2\pi l_s)^6 \int dC_7 \wedge ^{\star}dC_7 + \frac{1}{2\pi} \int C_7 \wedge \frac{F}{\sqrt{2}}\wedge \omega_1 +  \int C_7\wedge \omega_3\,, 
\label{ac7}
\ee 
is the ten-dimensional action for the Ramond-Ramond potential $C_7$. The kinetic term comes from the type IIA supergravity action; the second term is due to the flavor D8-brane, whose embedding, in the antipodal case, is described by a delta-like one form $\omega_1=\delta(y) dy$; the third term is due to the D6-brane source, whose embedding is described by the three form $\omega_3$.

Let us consider the equation of motion for $C_7$. Setting $^\star F_8 = (2\pi l_s)^{-6}\widetilde F_2$ we have, for D8-brane at $y=0$ and D6-brane at $x_3=z=y=0$,
\be
d\widetilde F_2 = \frac{F}{\sqrt{2}}\wedge \delta(y) dy + 2\pi \delta(z) \delta(y) \delta(x_3) dz\wedge dy\wedge d x_3\,,
\ee
where $F=dA$ is the $U(1)$ field strength on the D8-brane. 

We can formally solve the above equation (with $dC_1=0$) as 
\be
\widetilde F_2 = \frac{A}{\sqrt{2}}\wedge \delta(y) dy + 2\pi  \Theta(x_3) \delta(z) \delta(y) dz\wedge dy\,.
\ee
Now, the full action for $C_7$ turns out to be equivalent to\footnote{See e.g.~Appendix B in \cite{Bartolini:2016dbk}.}
\be\label{Seff}
S_{eff} = -\frac{1}{4\pi (2\pi l_s)^6}\int \widetilde F_2\wedge ^\star\widetilde F_2\,,
\ee
just as in the case with no explicit D6-brane source. With the same arguments as in \cite{Sakai:2004cn,Bartolini:2016dbk} we can thus obtain the four-dimensional effective Lagrangian for the $\eta'$ field  
as
\be\label{infinitelag}
{\cal L} = -\frac{\kappa}{\pi} (\partial_{\mu}\varphi)(\partial^{\mu}\varphi) + 2 c m  \cos\varphi -\frac{\chi_g}{2}\,\underset{k\in\mathbb{Z}}{\text{Min}}\,(\varphi-2\pi\Theta(x_3)+2\pi k)^2\,,
\ee
where we have also included the contribution of the flavor mass term ($m$ being the quark mass and $c$ a constant parameter) introduced in the string theory setup as in \cite{AK}. We have introduced $\chi_g$, that is the topological susceptibility of the unflavored gauge theory, which in units of $M_{KK}$ reads 
\be
\chi_g = \frac{\lambda^3}{4(3\pi)^6}\,.
\ee 
The Lagrangian above has been obtained 
%from the five-dimensional effective one on the D8-brane (reduced on $S^4$) (\ref{clacNf1})
by using the following ansatz for the gauge field
\be\label{Az}
A_z (z, x_3)= -\frac{\sqrt{2}}{\pi (1+z^2)}\varphi(x_3)\,,
\ee
and integrating over $z$. Hence
\be \label{fpi}
\int A_z dz \equiv -\sqrt{2}\varphi \equiv 2\frac{\eta'}{f_{\pi}}\,, \quad f_{\pi} = 2\sqrt{\frac{\kappa}{\pi}}\,.
\ee
From the Lagrangian (\ref{infinitelag}), we can derive the equation of motion for the field $\varphi = \varphi(x_3)$.  
We are going to search for a configuration interpolating from 0 for $x_3 \rightarrow - \infty$ to $2\pi$ for $x_3 \rightarrow + \infty$ and such that $\varphi(0)=\pi$.
In this case the potential is minimized for $k=0$ and the equation of motion
reads
\be\label{infeom}
\partial^{2}\varphi(x_3) - m_{WV}^{2}\left(\varphi(x_3) - 2\pi\Theta(x_3) \right) - m_\pi^2\sin(\varphi(x_3)) = 0,
\ee
where
\be
m_{WV}^2 \equiv \frac{\chi_g \pi}{2\kappa} = \frac{ \lambda^2}{27\pi^2 N}
\ee
is the Witten-Veneziano mass. Treating the D8-brane in the probe approximation amounts on taking $m_{WV}\ll 1$. The other parameter in the equation of motion above is
\be
m_{\pi}^{2} = \dfrac{4 cm}{f_{\pi}^{2}},
\ee
which represents the quark mass contribution to the $\eta'$ mass. When $N_f >1$ these become the standard parameters related to pions, as shown in \cite{AK}. As observed in \cite{Argurio:2018uup}, the parameter $m_{\pi}$ has to be small, $m_{\pi}\ll 1$, in order to neglect higher order corrections to the mass term of the WSS model.

\subsubsection*{Massless quark}

In the massless case $m=0$ the equation of motion for $\varphi$ (\ref{infeom}) has solution
\be\label{infinite}
\varphi(x_3) = \pi e^{m_{WV}\, x_3} - \pi e^{-m_{WV}\,x_3}\left(e^{m_{WV}\,x_3}-1\right)^2 \Theta(x_3)\,.
\ee
This solution has been derived in the presence of an explicit gluonic source.
Similar solutions without explicit sources, with regular or singular potentials, have a very long history, see e.g. \cite{Skyrme:1961vr,Gabadadze:2000vw,Forbes:2000et}.  

The thickness of the ``soft'' part of the wall (the one given by the $\eta'$ profile) is clearly linear in $1/m_{WV}$ (from (\ref{infinite})), so up to a numerical factor, 
\begin{equation}
\delta_{soft} \sim \frac{1}{m_{WV}} \sim \frac{\sqrt{N}}{\lambda M_{KK}}\,.
\end{equation}

\begin{figure}
\center
\includegraphics[scale=.65]{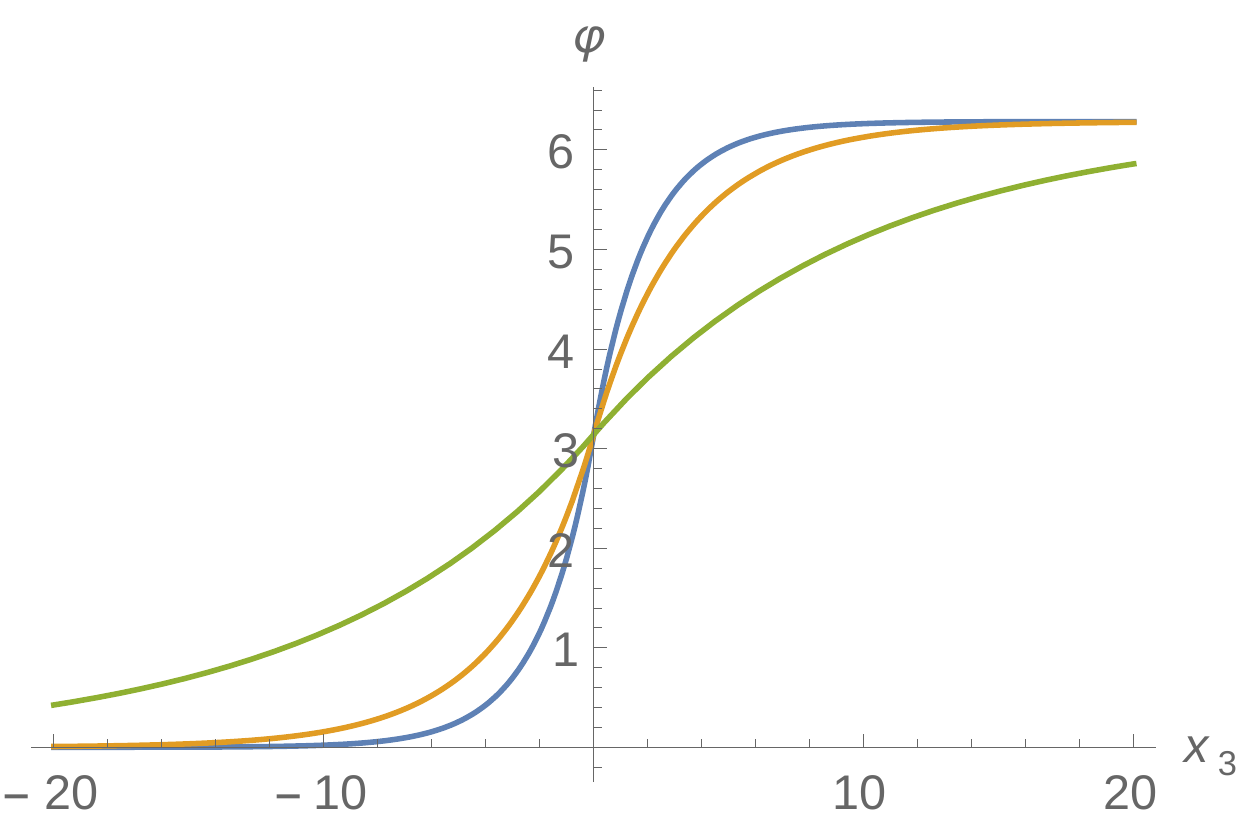}\includegraphics[scale=.65]{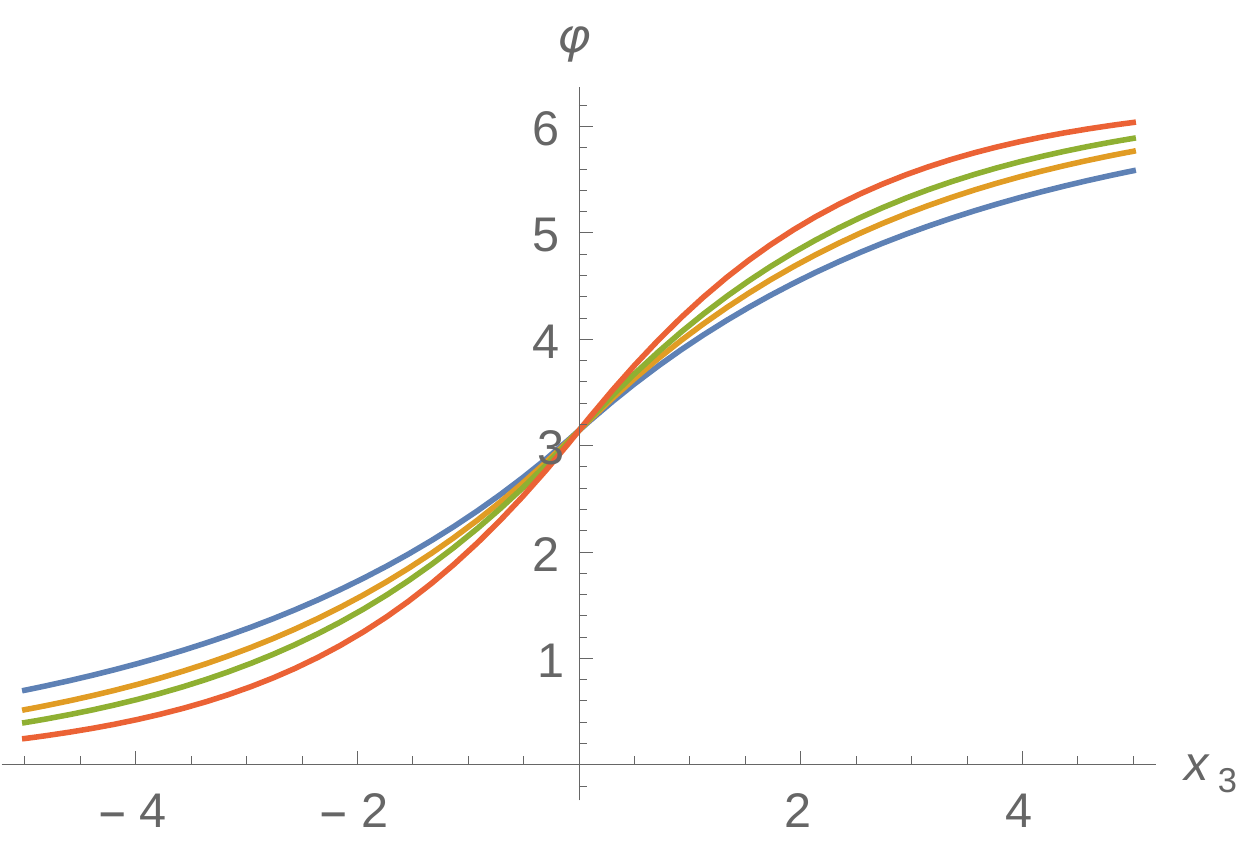}
\caption{Left: the profile of  $\varphi=-\sqrt{2}\eta'/f_{\pi}$ for $m_{WV}/M_{KK}=0.1, 0.3, 0.5$ (green, yellow and blue lines). Right: the profile of  $\varphi$  for $m_{WV}/M_{KK}=0.3$ and $m_{\pi}^2/M_{KK}^2=0, 0.05, 0.1, 0.2$ (blue, green, yellow and red lines).}
\label{fig1}
\end{figure}

The $\varphi \sim \eta'$ solution (\ref{infinite}) interpolates between zero (for $x_3\rightarrow-\infty$) and $2\pi$ ($x_3\rightarrow\infty$) and it takes the value $\varphi(0)=\pi$ in $x_3=0$. See figure \ref{fig1} (left panel).
Correspondingly, the Abelian field strength on the D8-brane reads
\be\label{infiniteF}
F_{x_3 z} = -\frac{\sqrt{2} m_{WV}}{1+z^2}\left(e^{m_{WV}\, x_3} - 2\Theta(x_3) \sinh (m_{WV}\, x_3)\right)\,,
\ee
where we have reinserted the $z$-dependent factor.  This satisfies 
\be
\int dx_3 dz F_{x_3 z} = -2 \sqrt{2}\pi\,,
\ee
hence, as anticipated in eq.~(\ref{cond1}) the $\eta'$ profile effectively induces a unit of $\overline{\rm D}6$-brane charge.

Due to the field strength (\ref{infiniteF}), the D8-brane has an excess of energy, providing the tension of the soft shell of the sheet. This can be deduced from the on-shell value of the action (\ref{clacNf1})
\begin{equation}\label{interm}
%\Delta E \simeq -\frac13 T_8 V(S^4) R^{3/2}u_0^{7/2} \int dt dx_1 dx_2 \int dz dx_3 (1+z^2)^{2/3} \frac{F^2}{2}\,.
S_f^{N_f=1} = -\frac{\kappa}{2}\int dt\, dx_1\, dx_2 \int dz\, dx_3 (1+z^2) F_{3z}^2\,.
\end{equation}
Inserting the solution (\ref{infiniteF}) and performing the integration in $z, x_3$ we get the tension
\begin{equation}\label{infinitetensionsoft}
T_{soft} =  \frac{1}{3^3 2^3 \pi^{2}}\lambda  N m_{WV} M_{KK}^2 = \frac{1}{3^{9/2} 2^3 \pi^{3}}\lambda^2  N^{1/2} M_{KK}^3\,.
\end{equation}
Note that considering the expression of $f_{\pi}$ in (\ref{fpi}), one has $T_{soft} \sim f_{\pi}^2 m_{\eta'}$.

\subsubsection*{Massive quark}

\begin{figure}
\center
\includegraphics[scale=.8]{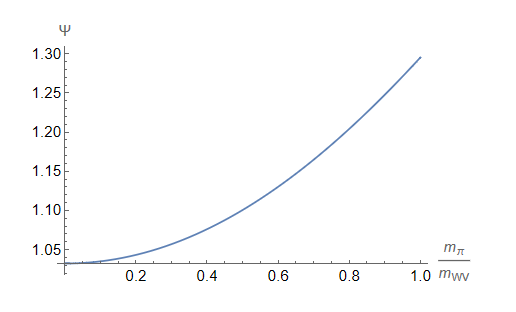}\caption{The function $\Psi(m_{\pi}/m_{WV})$ providing the quark-mass dependence of the soft part of the tension of the sheet.}
\label{figPsi}
\end{figure}
 
In the massive-quark case one can integrate numerically the equation derived from (\ref{infinitelag}) with $k=0$, with a similar behavior as in the massless case, see figure \ref{fig1} (right panel).
Also in this case the solution corresponds to a unit $\overline{\rm D}6$-brane charge.

The thickness of the soft part of the wall now depends on the quark mass $m$ (and hence on $m_{\pi}$) as well
and roughly behaves as (at small $\varphi$ this comes from (\ref{infinitelag}))
\begin{equation}
\delta_{soft} \sim \frac{1}{\sqrt{m_{WV}^2+m_{\pi}^2}}\,.
\end{equation}

In order to estimate the tension, we have to insert the numerical solution for $F_{x_3 z}$ in formula (\ref{interm}), which gives
\begin{equation}
T_{soft} =  \frac{1}{3^3 2^3 \pi^{2}}\lambda N m_{WV} \Psi(m_{\pi}/m_{WV}) M_{KK}^2 \,.
\end{equation}
The result is exactly as in (\ref{infinitetensionsoft}) but for the substitution
\be
m_{WV} \longrightarrow m_{WV}\Psi\left(\dfrac{m_{\pi}}{m_{WV}}\right),
\ee
where $\Psi(q)$ is a function, going as $\sqrt{1+q^2}$ at small $q$, which is reported in figure \ref{figPsi}.

\subsubsection*{Resume}

Table \ref{table} contains the parametric dependence of tension and thickness of the core and shell parts of the sheet discussed above.

\begin{table}[h!t]
\begin{center}
\begin{tabular}{|l|c|c|}
 \hline
 & Massless quark & Massive quark  \\
 \hline
Core tension $T_{hard}$  & $\lambda^2 N \Lambda^3 $ & $\lambda^2 N \Lambda^3 $ \\
Core thickness $\delta_{hard}$ & $\lambda^{-1/2}  \Lambda^{-1} $  & $\lambda^{-1/2}  \Lambda^{-1} $ \\
Shell tension $T_{soft}$ & $\lambda N m_{WV} \Lambda^2 \sim \lambda^2 N^{1/2} \Lambda^3$ & $\lambda N m_{WV} \Psi(m_{\pi}/m_{WV}) \Lambda^2$  \\
Shell thickness $\delta_{soft}$ & $m_{WV}^{-1} \sim \lambda^{-1} N^{1/2} \Lambda^{-1}$ & $(m_{WV}^2 + m_{\pi}^2)^{-1/2}$\\
\hline
\end{tabular}
\end{center}
\caption{Summary of parametric dependence of tension and thickness of the core and shell parts of the sheet. For the reader's convenience, we used the dynamical scale notation $\Lambda$ instead of $M_{KK}$.}
\label{table}
\end{table}

%%%%%%%%%%%%%%%%%%%%%%%%%%%%%%%%%%%%%%%%%%%%%%
\section{Semi-infinite sheet: the vortex string}
\label{secSemiInfinite}
\begin{figure}
\center
\includegraphics[scale=.9]{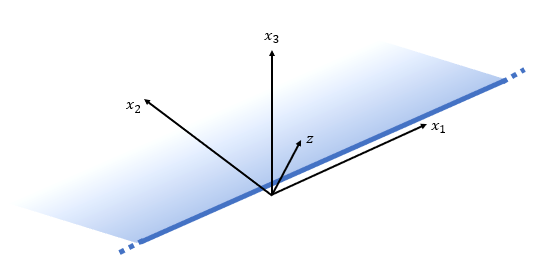}\caption{D6-brane dual to the semi-infinite sheet. It is taken to be extended along the $x_1$ axis and along the positive values of the $x_2$ axis. The directions $x_3$ and $z$ are orthogonal to the sheet. Every point is also extended in time and along other four dimensions which are wrapped on $S^4$.}
\label{figSemi}
\end{figure}

\setcounter{equation}{0}
In this section we describe the semi-infinite sheet, whose boundary is a straight infinite vortex string. 
The sheet is dual to a D6-brane which for convenience is placed at $x_3=z=y=0$ and extended along $x_2>0$.
The string is its boundary, that we take at $x_2=0$, attached to the D8-brane, see figure \ref{figSemi}.
The boundary is a magnetic source on the D8-brane worldvolume. %Since the D8-brane wraps the four-sphere, we can consider the reduced five-dimensional effective action (\ref{clac}), supplemented by the $C_7$ action terms accounting for the source
In this case the action terms (\ref{ac7}) take the form
\bea\label{action}
S_{C_7} & = & -\frac{1}{4\pi} (2\pi l_s)^6 \int dC_7 \wedge ^{\star}dC_7 \\\
&& + \frac{1}{2\pi} \int C_7 \wedge \left[\frac{F}{\sqrt{2}}\wedge \delta(y)dy + 2 \pi \Theta(x_2) \delta(x_3) dx_3 \wedge \delta(z) dz \wedge \delta(y) dy\right]\,. \nonumber
\eea
The corresponding equation for $^{\star}dC_7$,
\be\label{one}
d \widetilde F_2 = \frac{F}{\sqrt{2}}\wedge \delta(y)dy + 2 \pi \Theta(x_2) \delta(x_3) dx_3 \wedge \delta(z) dz \wedge \delta(y) dy\,,
\ee
provides the violation of the Bianchi identity due to the magnetic source (the boundary of the D6)
\begin{equation}\label{Bianchi}
d F = -2 \sqrt{2}\pi \delta(x_2) \delta(x_3)  \delta(z) dx_2 \wedge dx_3 \wedge dz\,.
\end{equation}

Thus, we are led to consider the problem of finding the fields generated by a uniform linear distribution of magnetic charge in five-dimensional electromagnetism with CS and $\widetilde F_2$ source terms on the curved background (\ref{curvedspace}).
%\begin{equation}\label{curvedspace}
%ds^2 = g(z) dx^{\mu}dx_{\mu} +  \frac{dz^2}{g(z)}\,, \qquad g(z)=(1+z^2)^{2/3}\,,
%\end{equation} 
%as can be deduced by rewriting (\ref{clac}) in general covariant form.
We have not found such type of solution, or part of it, in the literature, so we derive it in this section.
The strategy will be the following.
\begin{itemize}
\item[1.] We will first find a configuration ${F}^m$ satisfying the Bianchi equation (\ref{Bianchi}) and $d ^{\star}{F}^m=0$.
This corresponds to a kind of magnetic monopole-like solution for a source extended along the $x_1$ axis.
\item[2.] The complete field can be written as
\begin{equation}
{F} = {F}^m + d {A}^c\,,
\end{equation}
being $A^{c}$ the gauge connection associated to a closed field strength. If we consider a potential ${A}^m$ for ${F}^m$, we have
\begin{equation}
d{A}^m = {F}^m + {\cal S}\,,
\end{equation}
where ${\cal S}$ is a Dirac sheet (a Dirac string for an extended source).
A particular gauge choice of the Dirac sheet allows to solve equation (\ref{one}) as
\begin{equation}
\widetilde F_2 = \left(\frac{{A}^c+{A}^m}{\sqrt{2}}\right)\wedge \delta(y)dy + 2 \pi \Theta(x_3) \delta(z) dz \wedge \delta(y) dy\,.
\end{equation}
As in \cite{Sakai:2004cn} and Section \ref{sec:infinite}, the integral of this field along the $(z,y)$ cigar
\begin{equation}
\int_{cig} \widetilde F_2 = \int \left(\frac{{ A}^c_z+{A}_z^m}{\sqrt{2}}\right) dz + 2 \pi \Theta(x_3)\,,
\end{equation}
enters, through the supergravity term (\ref{Seff}), the equation for ${A}^c_z$.
If there are no CS terms (as it will be the case), and considering that $d ^{\star}{F}^m=0$, the equation reads schematically
\begin{equation}
d^{\star}d {A}^c_z = \int \left({A}^c_z+{A}_z^m\right) dz + 2 \pi \sqrt{2}\Theta(x_3)\,.
\end{equation}
Given that ${A}_z^m$ is known from step 1, we can solve this equation for ${A}^c_z$.
\item[3.] We can consider the charged case by switching on the components ${A}_{\pm}(x_{\pm})$ of the gauge field ($x^{\pm} = (x_1 \pm t)/2$).
One can easily realize that ${A}_{\pm}$ do not enter the previous equations, but on the contrary the field ${A}_z={A}^c_z + {A}^m_z$ generates a CS term for their equations, which schematically read
\begin{equation}
d^{\star}d {A}_{\pm} = {F}_{\pm x_2} \wedge {F}_{x_3 z}+ ...\,.
\end{equation}
The latter can be solved independently after steps 1 and 2 are completed, since the only unknowns are ${A}_{\pm}$.
\end{itemize}

In order to illustrate this procedure, in the next section we consider as a warm-up the flat space case.
Then, in section \ref{sec:semicurved} we will provide the solution on the curved space (\ref{curvedspace}).
The procedure does not depend on the explicit form of the metric function $g(z)$ and can in principle be applied to a generic metric of the form (\ref{curvedspace}).

%%%%%%%%%%%%%%%%%%%%%%%%%%%%%%%%%%%%%%%%%%%%%
\subsection{Solution in flat space}
\label{secFlat}

In this section we consider the case of five-dimensional flat spacetime, i.e.~$g(z)=1$.
In this case the configuration has spherical symmetry in the directions $x_2, x_3, z$ transverse to $t, x_1$ where the line source is placed, so we employ spherical coordinates
\begin{equation}
x_2 = \rho \cos\theta\,, \quad x_3 = \rho \sin\theta \cos\phi\,,\quad z = \rho \sin\theta \sin\phi\,.
\end{equation}
It is also convenient to parameterize the $(t,x_1)$-part of spacetime with light-cone coordinates
\be
x_\pm = \frac{x_1 \pm t}{2}\,,
\ee
such that
\be
ds^2 = 4 dx_+ dx_- +d\rho^2 + \rho^2 d\Omega_2\,,
\ee
with the two-sphere parameterized by angles $\theta, \phi$.

We consider an ansatz where:
\begin{itemize} 
\item all the fields just depend on the radial direction $\rho$;
\item the magnetic charge from equation (\ref{Bianchi}) is 
\bea
-2 \sqrt{2}\pi\delta^3(\rho) \rho^2 d\rho \wedge \omega_2\,,
\eea 
with $\omega_2=\sin\theta d\theta \wedge d\phi$ being the two-sphere volume form;
\item the field strength has just two components
\be
 F_{\theta \phi}\omega_2\,,\qquad F_{\pm \rho}dx_{\pm} \wedge d\rho \,,
\ee
where either $A_+$ or $A_-$ are turned on.
This implies that we have (only) one non-vanishing term in $F \wedge F$, entering the equation for $A_{\pm}$.
\end{itemize}
It is readily verified that the ansatz is consistent with the equations of motion and Bianchi identity.

%%%%%%%%%%%%%%%%%%%%%%%%%%%%%%%%%%%%%%%%%%%%%
\subsubsection{Step 1: monopole-like solution}

We start by searching for a solution ${F}^m$ to the Bianchi equation $d{F}^m = -2 \sqrt{2}\pi\delta^3(\rho) \rho^2 d\rho \wedge \omega_2\ $ and the equation of motion (without CS term, as explained above) $d^{\star} {F}^m=0$.

Considering translation invariance along $x_1$ and spherical symmetry, the problem is very similar to that of finding the field strength generated by the standard Dirac monopole in four dimensions.
Thus, the solution is quite similar
\begin{equation}\label{flatmonF}
{F}^m  = {F}^m_{\theta\phi}\omega_2 = -\frac{1}{2\rho^2} \rho^2 \sqrt{2} \omega_2\,,
\end{equation}
where $\omega_2 = \sin{\theta} d\theta \wedge d\phi$ and we have made explicit the singular behavior at the origin $\rho=0$.
It is straightforward to check that (\ref{flatmonF}) solves the equations of motion and that
\begin{equation}
\int_{M_3} d{F}^m = \int_{M_2} { F}^m = -2\sqrt{2}\pi\,,
\end{equation}
where $M_3, M_2$ are three- and two-dimensional closed surfaces containing the point $\rho=0$, $M_2$ being the boundary of $M_3$.

A standard way of writing a potential ${A}^m$ for this field is
\begin{equation}\label{infiniteflatA}
{A}^m = {A}^m_{\phi} d\phi = - \sqrt{2}\frac{(1-\cos\theta)}{2\rho\sin\theta}\rho\sin\theta d\phi\,.
\end{equation}
In this form, one has
\begin{equation}
d{A}^m = {F}^m + {\cal S}\,,
\end{equation}
where ${\cal S}$ is a Dirac sheet along $\theta=\pi$, i.e. extended along $x_1$ and the negative branch of $x_2$
\begin{equation}
{\cal S} = -2 \sqrt{2} \pi\Theta(-x_2)\delta(x_3)\delta(z) dx_3\wedge dz\,.
\end{equation}
As usual, one could choose the position of the Dirac sheet to be along any other direction, and glue the solutions up to gauge transformations.

The energy associated to the field strength (\ref{flatmonF}) is readily checked to be divergent, due to the small-$\rho$ behavior.
This can be interpreted as the usual divergence close to the charges (electric or magnetic) in standard electromagnetism.

The same field (\ref{flatmonF}) with sign changed represents the solution corresponding to an opposite sign of the charge distribution, $d{ F}^m = 2 \sqrt{2}\pi\, \delta^3(\rho) \rho^2 d\rho \wedge \omega_2$.
This corresponds to choosing an $\overline{\rm D}6$-brane instead of a D6-brane in the stringy setup.

%%%%%%%%%%%%%%%%%%%%%%%%%%%%%%%%%%%%%%%%%%%%%
\subsubsection{Step 2: solution for ${A}^c_z$}

In flat space this step is trivial, since there is no $(z,y)$-cigar over which we can integrate $\widetilde F_2$ with a finite result.
But our five-dimensional electromagnetic problem is well-defined without the $\widetilde F_2$ source term.
Without source, and since the CS term does not enter the equations for ${A}_z, { A}_{x_2}, {A}_{x_3}$, we can just consider (\ref{infiniteflatA}) as a solution of the equations, without the need to turn on further components in directions $x_2, x_3, z$. 

%%%%%%%%%%%%%%%%%%%%%%%%%%%%%%%%%%%%%%%%%%%%%
\subsubsection{Step 3: solution for ${A}_{\pm}$}

Even if it is not forced upon us by the equations, we can switch on the components ${A}_{\pm}(x_{\pm},\rho)$ of the gauge field.
The equations read
\be 
\frac{1}{g_5^2}d^{\star}{F} = N {F} \wedge {F}\,,
\ee
where the five-dimensional coupling constant is in this case $g_5^2 = 1/(8\pi^2\kappa)$.
Considering the solution for ${F}_{\theta\phi}$ (\ref{flatmonF}) in the CS term, 
the equations reduce to
\be\label{eqforApflat}
\partial_{\rho} (\rho^2 \partial_{\rho} {A}_{\pm})= \mp g_5^2 N \partial_{\rho} { A}_{\pm}\,,
\ee
with the solutions
\be\label{Frp}
\partial_{\rho} {A}_{\pm} = c(x_{\pm}) \frac{e^{\pm \frac{g_5^2N}{\rho} }}{\rho^2}\,.
\ee
Here we have included a possible arbitrary function $c(x_{\pm})$.\footnote{Note that the component ${F}_{+-}$ of the gauge field strength would mix the equations among each other, so we are keeping it off in this paper.}
Since $g_5^2 N \sim 1/(\lambda M_{KK})$, the exponential behavior of the solution, due to the CS term, sets in for radii smaller than $\rho \sim 1/(\lambda M_{KK})$, that is very close to the source (since $\lambda \gg 1$). 

Now we can note two interesting facts.
First, only one of the two modes, ${A}_{-}$,\footnote{The fact that $A_-$ is normalizable depends on the sign choice of the magnetic source, i.e.~on whether we consider a D6 or an $\overline{\rm D}6$-brane in the string setup.} is normalizable.
Second, if we just keep the normalizable mode, i.e.~as long as we do not switch on both components ${F}_{\rho -}$ and ${F}_{\rho +}$ in (\ref{Frp}), it has zero energy: it is tempting to read this as an indication that the mode $c(x_{\pm})$ has zero potential.
These properties seem to suggest that the functions $c(x_{\pm})$ could be related to the chiral edge modes expected to be present on the boundary of the sheet: they are only left or right moving and have zero Hamiltonian.

%%%%%%%%%%%%%%%%%%%%%%%%%%%%%%%%%%%%%%%%%%%%%%
\subsection{Solution on curved space}
\label{sec:semicurved}

In this section we report on the main result of this paper, that is the solution for the semi-infinite sheet on the curved background (\ref{curvedspace}).
For simplicity, we just consider the massless-quark case.

The boundary of the sheet extends along $t, x_1$, while the sheet itself is placed at $z=x_3=0$ and extends along $x_2 \geq 0$ up to the boundary at $x_2=0$, see figure \ref{figSemi}.
The configuration without the sheet has translational invariance along $x_1$ and axial symmetry around the boundary, so we can use polar coordinates\footnote{The curvature along $z$ breaks the spherical symmetry we had in flat space. The difference between the angular variable $\varphi$ and the normalized $\eta'/f_{\pi}$ field should be hopefully clear from the context.}
\begin{equation}
x_2=r \cos{\varphi}\,, \qquad  x_3=r \sin{\varphi}\,,
\end{equation}
in terms of which the background metric reads
\be
ds^2 =g(z) (4dx_+dx_- + dr^2 + r^2 d\varphi^2) +  \frac{dz^2}{g(z)}\,.
\label{curvedspaceh}
\ee
The ansatz we put forward is such that:
\begin{itemize} 
\item all the fields are independent on $x_\pm$ (eventually apart from the field ${A}_{\pm}$ which can depend on $x_{\pm}$);
\item the field strength has the components
\be
F_{x_2 z}\,,\quad F_{x_3 z}\,,\quad F_{x_2 x_3}\,,\quad F_{\pm z} \,,\quad F_{\pm x_2}\,,\quad F_{\pm x_3}\,.
\ee
Thus, the non-vanishing CS terms have components
\be
F_{x_2 z}F_{\pm x_3} \,, \quad F_{x_3 z}F_{\pm x_2} \,, \quad  F_{x_2 x_3}F_{\pm z} \,, 
\ee
and all enter just the equation for $A_{\pm}$.
\end{itemize}
The ansatz is consistent with the equations of motion and Bianchi identity (\ref{Bianchi}).
We proceed to solve them in three steps as explained above.

%%%%%%%%%%%%%%%%%%%%%%%%%%%%%%%%
\subsubsection{The vortex string}

The goal of this section is to find a solution $F^m$ of the Bianchi equation (\ref{Bianchi}) such that $d^{\star}F^m=0$ on the background (\ref{curvedspaceh}).
In this sub-section the only non-trivial fields will be $F_{x_2 z}, F_{x_3 z}, F_{x_2 x_3}$.

The configuration is less symmetric than in flat space, so we are not able to write a compact solution.
But we can expand the fields in a basis of eigenfunctions depending only on $z$ and finding the corresponding modes.

We start by writing an ansatz which solves automatically the equation of motion $d^{\star}F^m=0$, which reads
\be
F^m_{x_2 z} = - g(z)^{-3/2} \partial_{x_3}H\,, \quad F^m_{x_3 z}=g(z)^{-3/2} \partial_{x_2}H\,, \quad F^m_{x_2 x_3}=g(z)^{1/2} \partial_{z}H\,,
\label{ansazz}
\ee
with the function $g(z)$ given in (\ref{curvedspace}).
This form can also be written as
\be \label{semiF}
F^m=   d\varphi \wedge [g(z)^{-3/2} r \partial_r H dz -  g(z)^{1/2} r \partial_z H dr ]\,.
\ee

The function $H=H(x_2,x_3,z)$ is determined by the Bianchi equation (\ref{Bianchi}), which reads
\be \label{bianchih}
g(z)^{-3/2} (\partial_{x_2}^2 + \partial_{x_3}^2) H +\partial_z (g^{1/2} \partial_z H) = -2\sqrt{2}\pi \delta(x_2)\delta(x_3)\delta(z-Z)\,,
\ee
where $Z$ is a (possibly pseudo)-modulus for the position of the string in the $z$ direction. Here we consider just the classical configuration taking $Z=0$ at the end, but we should remind that, eventually, $Z$ should become a time dependent operator at the quantum level.
Taking inspiration from similar equations in \cite{SS-barioni}, we search for a solution in the form
\be \label{ansH}
H(x_2,x_3,z) = \sum_{n=0}^{\infty} Y_n(r) \zeta_n(z)\zeta_n(Z)\,,
\ee
where we have set to zero, without loss of generality, the modulus\footnote{$R = \sqrt{X_{2}^{2} + X_{3}^{2}}$, where the moduli $X_{2}$ and $X_{3}$ together with the pseudo-modulus $Z$ represent the center of mass collective coordinates of the vortex string.} $R$ in the argument of the $Y_n(r-R)$.
Equation (\ref{bianchih}) is equivalent to the infinite set of equations
\begin{eqnarray}
&& (\partial_{x_2}^2 + \partial_{x_3}^2 - \lambda_n) Y_n =-2\sqrt{2}\pi \delta(x_2)\delta(x_3)\,,\label{eqY}\\
&& g(z)^{3/2}\partial_z (g^{1/2} \partial_z \zeta_n)+\lambda_n \zeta_n = 0\,,\label{eqPsi}
\end{eqnarray}
supported by the completeness condition
\be \label{complete}
\sum_{n=0}^{\infty} g(z)^{-3/2} \zeta_n(z) \zeta_n(Z) = \delta(z-Z)\,,
\ee
satisfied by the functions 
$\zeta_n(z)$ with orthonormality condition given by
\be 
\int dz g(z)^{-3/2} \zeta_n(z)\zeta_m(z) = \delta_{nm}\,.
\ee

As long as $\lambda_n \neq 0$, the first set of equations (\ref{eqY}) has solutions in terms of the modified Bessel function of order zero of the second kind\footnote{Remember that we are working in units where $M_{KK}=1$. One can obtain the correct dimensionalities by reinserting the appropriate powers of $M_{KK}$ in these formulas.}
\be \label{solY}
Y_n(r) = \sqrt{2} K_0(\sqrt{\lambda_n}r)\,, \qquad n>0 \,.
\ee
They behave as logarithms at small $r$, while fall off exponentially fast at large $r$.
For $\lambda_{0} \equiv 0$ the solution of (\ref{eqY}) is just the logarithm
\be \label{solY0}
Y_0(r) = -\sqrt{2} \log{(r)} \,, \qquad n=0\,.
\ee
We have chosen to put to zero the constant solution, since the ansatz (\ref{ansazz}) only contains derivatives of $H$. 

We have numerically solved the second set of equations (\ref{eqPsi}), determining the eigenvalues $\lambda_n$, by requiring that the derivatives of the functions $\zeta_n(z)$ vanish at $z \rightarrow \pm \infty$ for $n>0$, so that they are orthogonal to the even part of the zero-mode, which is just a constant, $\zeta_0=1/\sqrt{\pi}$.
The masses of the first even modes computed from (\ref{eqPsi}) are then
\be
\lambda_{n} = 0, 1.6, 4.6, 9.2, 15.2, 22.8, 31.8, \dots
\ee
We have checked that the numerical solutions form a complete set according to (\ref{complete}).

\begin{figure}
\center
\includegraphics[scale=.6]{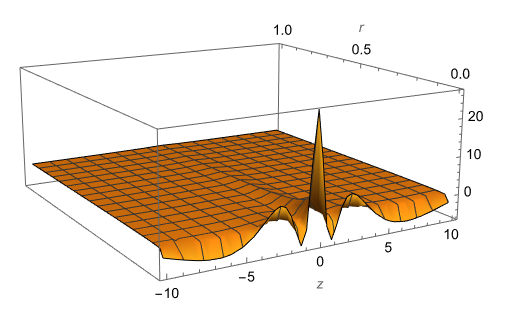}
\includegraphics[scale=.6]{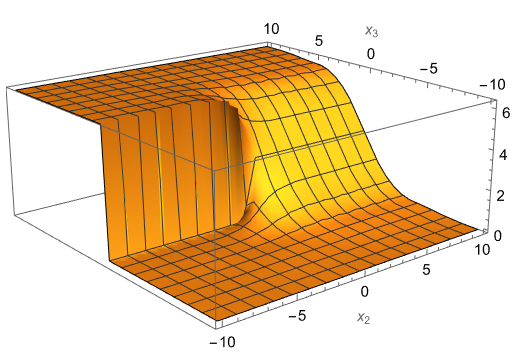}
\caption{Left: the function $H(r,z)$ obtained from the first six (even) modes in (\ref{ansH}). Right: The profile of $-\frac{\sqrt{2}}{f_{\pi}}\eta'$ for $m_{WV}/M_{KK}=0.3$.}
\label{figH}
\end{figure}

Notice that since the functions $\zeta_n$ have definite parity, so that $\zeta_n(Z=0)=0$ for odd parity, classically the function $H$ is even in $z$. 
A plot of the resulting solution for $H$ is reported in figure \ref{figH} (left panel).

It can be checked numerically that with such solution in (\ref{semiF}) one has the correct magnetic charge in (\ref{Bianchi})
\begin{equation}
\int_{M_3} dF^m = \int_{M_2} F^m = -2\sqrt{2}\pi\,,
\end{equation}
where $M_3, M_2$ are three- and two-dimensional closed surfaces containing the point $r=z=0$, $M_2$ being the boundary of $M_3$.\footnote{We used spheres of different radii to perform the check.} 

We can choose to express the solution in terms of the potentials
\begin{eqnarray}\label{gauge}
A^m_z = g(z)^{-3/2} r \varphi \partial_{r}H\,, \qquad A^m_r = -g(z)^{1/2} r \varphi \partial_{z}H\,.
\end{eqnarray}
This corresponds to choosing the Dirac sheet ${\cal S}$, defined by
\be 
dA^m = F^m + {\cal S}\,,
\ee
to be placed along the negative $x_2$ direction
\be \label{Dirac}
{\cal S} = -2\sqrt{2} \pi \Theta(-x_2)\delta(x_3)\delta(z) dx_3 \wedge dz\,.
\ee

We can see explicitly the stringy nature of the solution by plugging formulas (\ref{semiF}), (\ref{ansH}) in the action (\ref{clacNf1}) and integrating in $z$, obtaining the four-dimensional effective action for the vortex string
\be \label{actionvortex}
S^{N_f = 1}_f = -\frac12 \sum_{n=0}^{\infty} \int d^4x \left[(\partial_r \widetilde Y_n(r,\varphi))^2 + \lambda_n  (\widetilde Y_n(r,\varphi))^2 \right]\,,
\ee
where
\be 
\widetilde Y_n \equiv \sqrt{\kappa} \zeta_n(0)Y_n\,.
\ee
As we can see, we have an infinite tower of massive modes depending only on the two coordinates $x_2, x_3$ transverse to the string (with dynamics only in the radial direction), plus a massless zero-mode.

Putting the Lagrangian density in (\ref{actionvortex}) on-shell on the solutions (\ref{solY}) and integrating in $r, \varphi$, one can derive the tension of the vortex string (re-inserting the $M_{KK}$ factors) 
\be \label{divergent}
T_v = M_{KK}^2 \frac{\lambda N}{108 \pi^3}  \int_0^{\infty} dr \Biggl\{\frac{1}{r} +\pi r \sum_{n=1}^{\infty} \zeta_n(0)^2 \lambda_n  \left[K_0^2(\sqrt{\lambda_n} r) + K_1^2(\sqrt{\lambda_n} r) \right] \Biggr\}\,.
\ee
As expected \cite{Gabadadze:2000vw,Forbes:2000et,Gabadadze:2002ff,Komargodski:2018odf}, this scales linearly with $N$ and is divergent at the location $r=0$ of the string.
One has to remember that the solution is derived under the hypothesis that the field strength is small in order to consider only the first term in the expansion of the square-root of the DBI action.
Thus, the solution we found cannot be trusted close to the source, where the divergence is expected to be avoided in the full DBI setup.

On the other hand, we do know that the core of the string is made by the boundary of the D6 on the D8-brane, and that it has the width (\ref{thickhard}) 
\be \label{deltavhard}
\delta_{v,hard} \sim  \frac{1}{\lambda^{1/2}\Lambda}\,.
\ee
This can thus be reasonably used to cut-off the integral (\ref{divergent}) at small $r$.

There is also a divergence of (\ref{divergent}) at large distances, due to the zero mode.
It is similar to the one of the classical global $U(1)$ string.
In that context, it is expected to be cut-off by the presence of other strings or other physical effects.\footnote{For example it can be regulated by the curvature of spacetime in a cosmological context.}
Calling $R_{c}$ the long distance cut-off, one obtains the estimate for the tension of the ``soft'' shell of the string as
\be 
T_{v,soft} \sim  \Lambda^2 \lambda N \log{(R_c/\delta_{v,hard})} \sim \Lambda^2 \lambda N \log{(\lambda^{1/2}R_c\Lambda)}\,, 
\ee
while the ``hard'' core of the strings, coming from the D6-brane, has tension
\be 
T_{v,hard} \sim \Lambda^2 \lambda^{3/2} N\,.
\ee  
This follows by considering the boundary of the sheet to have radius $\delta_{v,hard}$ (\ref{deltavhard}), which is a reasonable definition in the antipodal D8-brane configuration we are considering. In the non-antipodal cases, the power of $\lambda$ is expected to be equal to two.

Finally, the logarithmic behavior of the zero mode means that the soft shell of the string is very spread in space, so that it is not meaningful to identify a thickness different from the above-mentioned large-distance cut-off and 
\be 
\delta_{v,soft} \sim R_c\,.
\ee
The above estimates are summarized in table \ref{tablev}.
\begin{table}[h!t]
\begin{center}
\begin{tabular}{|l|c|}
 \hline
Core tension $T_{v,hard}$  & $\lambda^{3/2} N \Lambda^2 $ \\
Core thickness $\delta_{v,hard}$ & $\lambda^{-1/2}  \Lambda^{-1} $  \\
Shell tension $T_{v,soft}$ & $ \lambda  \log{(\lambda^{1/2}R_c\Lambda)} N \Lambda^2$ \\
Shell thickness $\delta_{v,soft}$ & $R_c$\\
\hline
\end{tabular}
\end{center}
\caption{Summary of parametric dependence of tension and thickness of the core and shell parts of the vortex loop (string). For the reader's convenience, we used the dynamical scale notation $\Lambda$ instead of $M_{KK}$. $R_c$ is a large-distance cut-off.}
\label{tablev}
\end{table}

\subsubsection{The sheet}

In this section we work out the solution of the equation of motion for the closed part $A^c_z$ of $A_z$ including the source from $\widetilde F_2$.
%The solution corresponds to the sheet wall.

Let us begin by computing the term in $\widetilde F_2$.
Consider equation (\ref{one}) for $d\widetilde F_2$ with the gauge field
\be 
F^m + dA^c = dA^m + {\cal S} + dA^c\,.
\ee
With the choice of Dirac sheet as in (\ref{Dirac}) we can solve (\ref{one}) as
\be 
\widetilde F_2 = \left[ \frac{1}{\sqrt{2}}(A^m + A^c) +2\pi \Theta(x_3)\delta(z)  dz \right] \wedge \delta(y) dy \,.
\ee
Its integral on the cigar gives
\be \label{flux}
\int_{cig} \widetilde F_2 = \frac{1}{\sqrt{2}}\int (A^m_z + A^c_z)dz + 2\pi \Theta(x_3)\,.
\ee
As before, this gives
 a source term in the equation for $A^c_z$ of the form
\be 
-\frac{\chi_g}{2} \left[ \frac{1}{\sqrt{2}}\int (A^m_z + A^c_z)dz + 2\pi \Theta(x_3) \right]^2\,.
\ee 
Note that we know the explicit form of $A^m_z$, and so its integral, from (\ref{gauge}).

The $z$-dependence of the field $A^c_z$ can be dealt with the ansatz (\ref{Az})
\be  
A^c_z =-\frac{\sqrt{2}}{\pi g(z)^{3/2}}\phi(x_2,x_3)\,.
\ee
Its equation of motion reduces to an equation for $\phi(x_2,x_3)$
\be 
(\partial_{x_2}^2+\partial_{x_3}^2) \phi - m_{WV}^2 \left(\phi -\frac{1}{\sqrt{2}}\int A^m_z dz  -2\pi \Theta(x_3)  \right)=0\,.
\ee
It is straightforward to integrate numerically this equation.
The sum of the solution and the integral of $A^m_z$ gives the profile of the $\eta'$ field
\be 
- \frac{\sqrt{2}}{f_{\pi}}\eta' =  \left(\phi(x_2,x_3) -\frac{1}{\sqrt{2}} \int A^m_z dz  \right)\,.
\ee
We impose the boundary conditions $\eta'(x_2,-\infty)=0$ (trivial vacuum on one side of the sheet), $\eta'(\infty,x_3)=\eta'_{\infty}(x_3)$, where $\eta'_{\infty}(x_3)$ is the profile found in section \ref{secinf} for the infinite sheet.
We plot this field in figure \ref{figH} (right panel). 
While for $x_2>0$ the profile resembles the one of the infinite sheet by construction, for $x_2<0$ it is visible the discontinuity due to the Dirac sheet, enforcing the monodromy of the $\eta'$ field around the D6-brane wall (whose boundary is at $x_2=x_3=0$ in Minkowski). 

As usual, the solution induces a D6-brane charge on the D8-brane, which we can calculate in terms of the flux of $F_{x_3 z}$, obtaining
\be
\int dx_3 dz F_{x_3 z} = \int dz A_z|^{x_3=+\infty}_{x_3=-\infty} - \int dx_3 dz {\cal S}_{x_3 z} = -2 \sqrt{2}\pi \Theta(x_2)\,,
\ee
where ${\cal S}$ is the Dirac sheet in formula (\ref{Dirac}).  
The result is that there is a unit $\overline{\rm D}6$-brane charge induced only for $x_2>0$, where the sheet is located. Once again, the condition (\ref{cond1}) accounting for the cancelation of the total D6-brane charge in the setup is satisfied. 

The parametric dependence of the tension and thickness of the sheet are the same as in the infinite sheet case, see table \ref{table}.

The effective action for the system of the vortex string and the sheet can be derived from the actions  (\ref{clacNf1}), (\ref{Seff}) considering that the complete field strength reads
\begin{eqnarray}
&& F_{x_2 x_3} = g(z)^{1/2} \partial_z H\,, \\ 
&& F_{x_2 z} = - g(z)^{-3/2} \left[\partial_{x_3} H + \frac{\sqrt{2}}{\pi}\partial_{x_2} \phi \right]\,, \\
&& F_{x_3 z} = g(z)^{-3/2} \left[\partial_{x_2} H - \frac{\sqrt{2}}{\pi}\partial_{x_3} \phi \right]\,.
\end{eqnarray}
By means of formula (\ref{ansH}) and integrating in $z$ the result can be derived to be 
\begin{eqnarray} \label{effectiveaction}
S &=& -\frac12 \int d^4x \Biggl\{ \sum_{n=0}^{\infty}  \left[ (\partial_r \widetilde Y_n)^2 + \lambda_n  (\widetilde Y_n)^2 \right] + \left[(\partial_{x_2} \widetilde\phi)^2 + (\partial_{x_3} \widetilde \phi)^2 \right] \nonumber \\
&& \qquad \qquad \qquad + \chi_g \left[\sqrt{\frac{\pi}{2\kappa}} \widetilde \phi -\frac{1}{\sqrt{2}}\int A^m_z dz  -2\pi \Theta(x_3)   \right]^2 \Biggr\}\,,
\end{eqnarray}
where
\be 
\widetilde Y_n \equiv \sqrt{\kappa} \zeta_n(0)Y_n\,, \qquad \widetilde\phi \equiv \sqrt{\frac{2\kappa}{\pi}} \phi\,,
\ee
and $\int A_z^m dz$ can be given as a series of the $\widetilde Y_n$ using formulas (\ref{gauge}), (\ref{ansH}).
The action (\ref{effectiveaction}) is that for a string configuration (the part in $\widetilde Y_n$) and a membrane (the part in $\widetilde \phi$) with a non-trivial potential.

This action must be used with caution. 
In fact, while it gives the correct equation of motion for $\widetilde \phi$, it does not provide the correct equations for the $\widetilde Y_n$.
This is because of the coupling term between the $\widetilde Y_n$ in $\int A_z^m$ and the $\widetilde \phi$ in the second line of (\ref{effectiveaction}), which must be dropped by hands from the equations for the $\widetilde Y_n$.
Ultimately, the reason for this resides in the fact that the $\widetilde Y_n$ and $\widetilde \phi$ are both modes of the same gauge field in five dimensions.
We chose to solve the equation of motion of the latter by enforcing the extra constraint $d^{\star} F^m =0$ on the monopole-like part of the solution (which is written in terms of the $\widetilde Y_n$).
The latter constraint is not included in (\ref{effectiveaction}) and amounts to dropping the coupling term between the $\widetilde Y_n$ and the $\widetilde \phi$ in the equations of motion of the $\widetilde Y_n$, as stated above.%\footnote{This could be possibly enforced by a Lagrange multiplier.}

In any case, we made an extra check that the configuration we found in two-steps, first solving for $F^m$ enforcing the constraint $d^{\star} F^m =0$, and then solving for the rest of $F$, satisfies the complete equation of motion for the five-dimensional gauge field.

\subsubsection{The charged mode}
 
In this section we solve for the components $A_{\pm}$ of the D8-brane gauge field.
Their equations of motion reduce to
\begin{align}
\label{eqforAp}
&\partial_{z}\left(g^{3/2}(z)\partial_{z}{A}_{\pm}\right) + \dfrac{1}{\sqrt{g(z)}}\left(\partial_{x_3}^{2} + \partial_{x_2}^{2}\right){A}_{\pm} 
\pm \dfrac{N}{16\pi^{2}\kappa}\Bigl[\sqrt{g(z)}\partial_{z}{A}_{\pm}\partial_{z}H + \,\, \nonumber\\ 
& + \dfrac{\partial_{x_2}{A}_{\pm}}{g^{3/2}(z)}\left(\partial_{x_2} H - \frac{\sqrt{2}}{\pi}\partial_{x_3} \phi \right) + \dfrac{\partial_{x_3}{A}_{\pm}}{g^{3/2}(z)}\left(\partial_{x_3} H + \frac{\sqrt{2}}{\pi}\partial_{x_2} \phi \right)\Biggr] = 0 \,.
\end{align}
Since the only unknown in these equations are $A_{\pm}$, one can find them either numerically or employing the basis $\zeta_n(z)$ to reduce them to a set of equations in just the $x_2, x_3$ variables.
In any case, the solutions can be multiplied by a generic function $c(x_{\pm})$, as in the flat space case.
The associated energy is zero as long as there is just one of the two components $F_{\pm M}$ turned on ($M=x_\mp,x_2,x_3,z$).
Again, this mode could be related to the chiral edge mode expected on the border of the sheet.

In fact, one can argue without solving equation (\ref{eqforAp}) that only one of the two modes $A_{\pm}$ is normalizable.
The reason is that the equation simplifies drastically close to the string at $z \sim x_2 \sim x_3 \sim 0$.
It can be checked that it reduces to the flat space one (\ref{eqforApflat}), with the two solutions (\ref{Frp}), only one of which ($A_-$) is normalizable at the origin.
Thus, there is only a left moving or a right moving normalizable mode.

One can derive an explicit numeric solution for this mode given particular boundary conditions. We leave to a future study the search for a more general solution.
The analysis of the equation in the asymptotic regions of large $z, x_2, x_3$ shows that one can impose the solution to go to constant values.
It is natural to choose a zero constant at large $|z|$.
Another boundary condition can be imposed at $z=0$, where equation (\ref{eqforAp}) can be integrated numerically in $x_2, x_3$.
In this case one can impose that close to the origin the solution is the flat space one (a primitive of equation (\ref{Frp}) at $z=0$), and that at large $|x_2|, |x_3|$ it goes to a constant.\footnote{This condition is the same as in flat space as well. The difference between the flat space case and the curved one (\ref{eqforAp}) at $z=0$ is encoded in the two terms $\partial_{x_2,x_3} \phi$. These are limited and multiplied by derivatives $\partial_{x_2,x_3} A_\pm$, so a constant boundary condition is still the most natural one.} 

\begin{figure}
\center
\includegraphics[scale=.6]{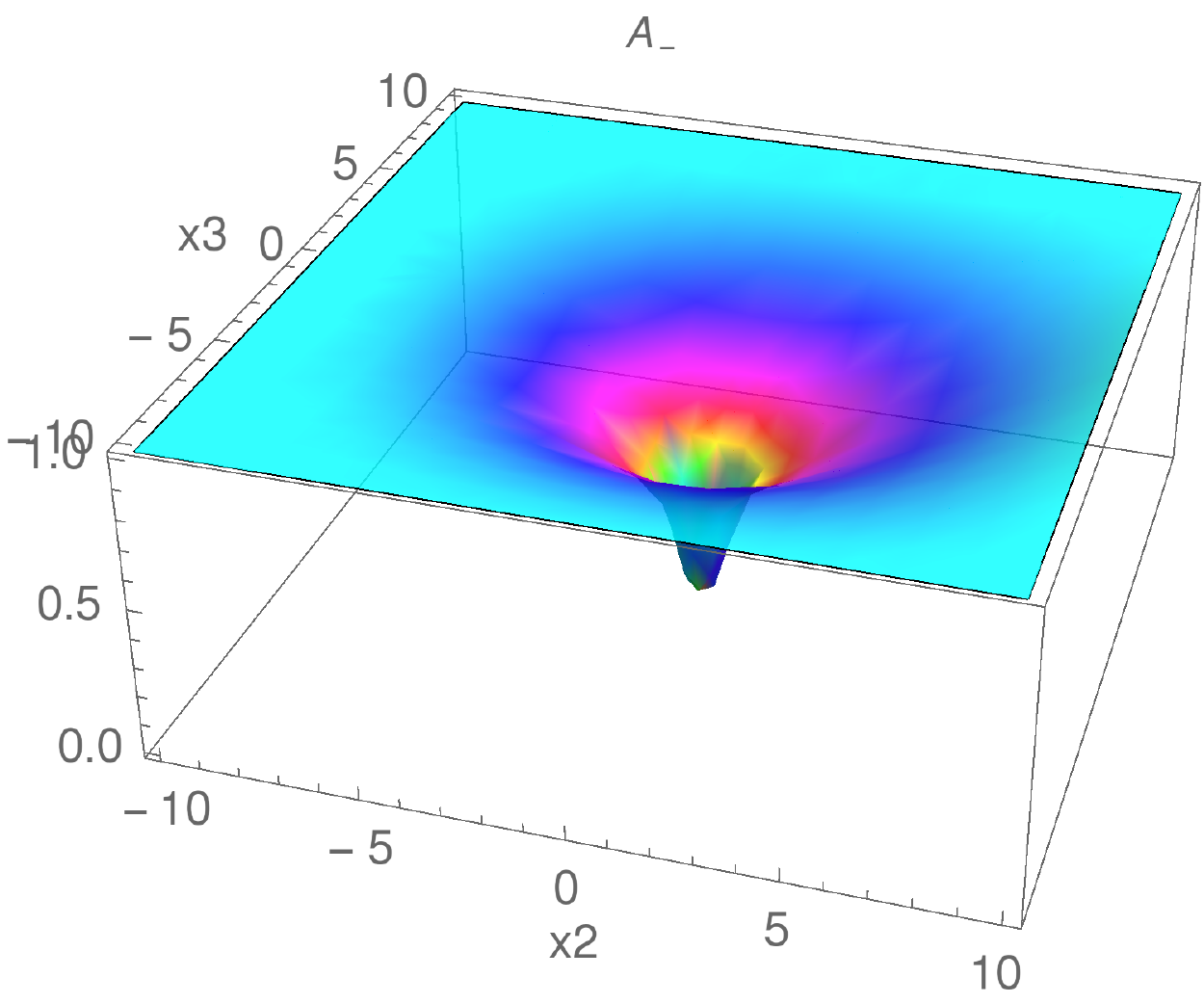}\includegraphics[scale=.6]{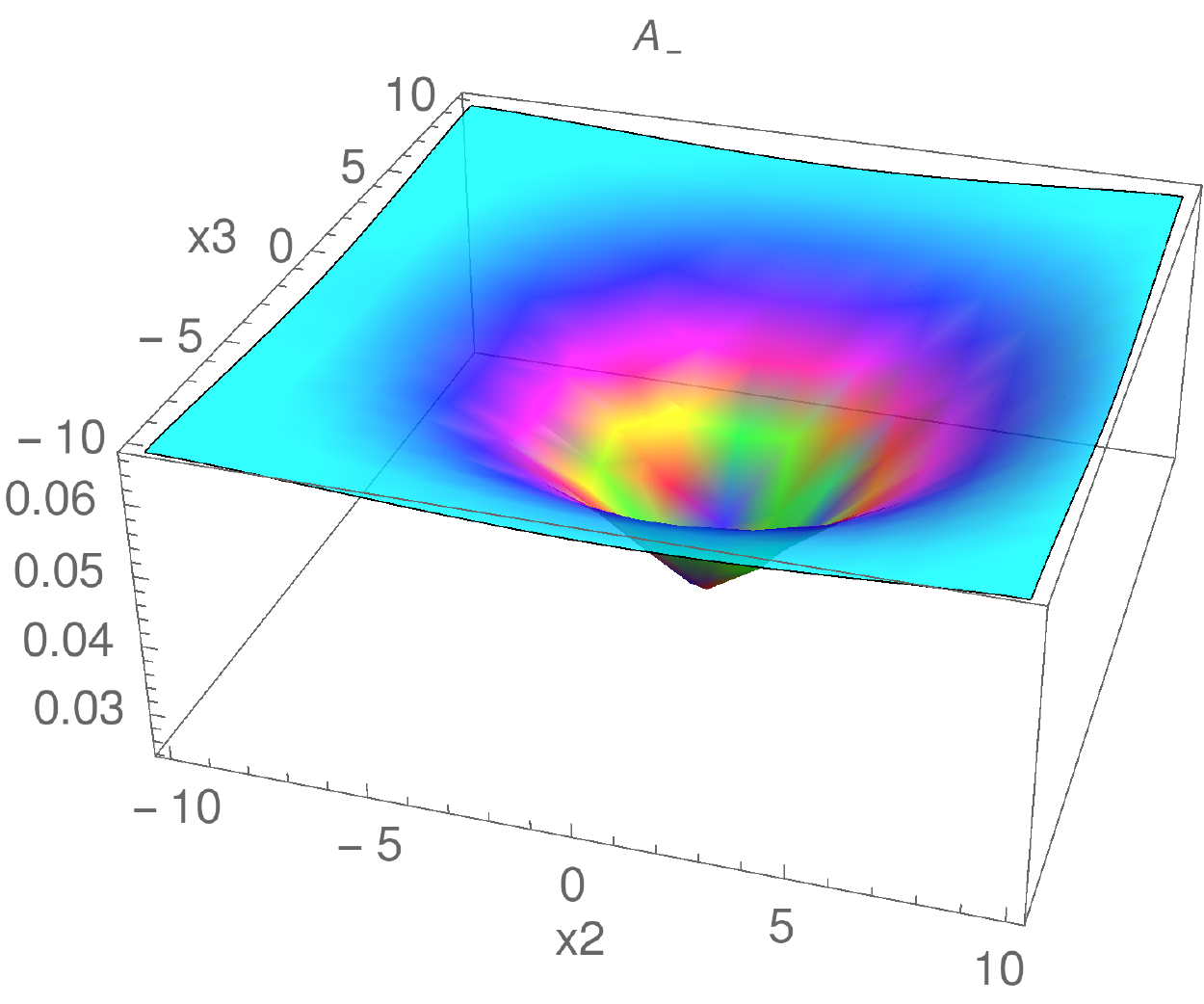}
\caption{Slices of the solution for $A_-$ at $z=0$ (left) and $z=5$ (right). For these plots we fixed $N/16\pi^2 \kappa =1$ and $A_-(x_2,x_3,0)=1$ at large $|x_2|, |x_3|$. The $z=0$ slice is almost the same as in flat space. As $z$ increases, the solution reduces in amplitude and eventually goes to zero at large $z$ (as dictated by the boundary conditions).}
\label{figAp}
\end{figure}

With these boundary conditions, one can integrate numerically equation (\ref{eqforAp}).
Plots representing two slices of the solution at $z=0$ (the boundary condition) and at a fixed $z>0$ are reported in figure \ref{figAp}. 

%%%%%%%%%%%%%%%%%%%%%%%%%%%%%%%%%%%%%%%%%%%%%%
\section{Summary and future developments}
\label{secConclusions}

%Let us summarize the content of these notes.
In this paper we have put forward a proposal for the string dual, in the context of the WSS model of holographic QCD, of the quantum Hall droplet, or sheet, that should describe one-flavor baryons at low energies \cite{Komargodski:2018odf}.
This object consists of a D6-brane wrapped on the four-sphere of the ten-dimensional background (much like the dual of the domain wall at $\theta=\pi$), together with the non-trivial D8-brane gauge field sourced by the D6-brane.
The D6-brane constitutes the hard, gluonic core of the sheet, responsible for the non-analyticity of the $\eta'$ low-energy action.
The D8-brane gauge field provides the soft, mesonic (in particular $\eta'$) part of the sheet.

Concerning QCD, the holographic construction highlights the possibility of the scenario where the sheet is realized in a Higgs phase.
The latter corresponds to the melting of the D6 in the D8 world-volume.
In this paper we have instead considered the scenario where the D6 are not melted, employing anyway the simplified setting of antipodal D8 embedding.  

After providing a few evidences for the proposal, e.g.~the emergence of the expected CS theory on the world-volume of the sheet, we have constructed two explicit configurations.
The first one corresponds to an infinitely extended sheet without boundaries.
The profile of the $\eta'$, interpolating from 0 to $2\pi$ on the two sides of the sheet, has been derived.
The holographic construction allows to calculate precisely the tension and thickness of the sheet, both in the massless and massive-quark cases.

The second configuration we have considered is semi-infinite, with a straight infinite boundary (where the D6 terminate on the D8).
The boundary constitutes a continuous distribution of magnetic charge for the five-dimensional Maxwell-CS theory (on curved space) on the D8-brane.
We solved for this interesting problem both in flat space, where an analytic form for the gauge field can be provided, and on the curved Witten background.
In the latter case, we have provided a semi-analytic solution.
We have derived a four-dimensional effective action for the fluctuating modes and calculated the tension and thickness of the hard and soft part of the string-like boundary of the sheet.

Finally, we have observed that one can switch on a chiral component of the gauge field on the boundary having zero energy, which is possibly connected to the chiral edge mode one could consider on the boundary of the Hall droplet.     

We list a few directions for possible developments of the work presented in this paper.
\begin{itemize}
\item A few issues that we left open include 
a better understanding of the $c_{\pm}(x_{\pm})$ mode in relation to the chiral edge mode;  the exploration of other choices of the Dirac sheet and the explicit construction of a Wu-Yang monopole-like solution without Dirac sheet; the possibility of the existence of a more general solution relaxing the constraint $d^{\star} F^m=0$. %(or would it be just another gauge choice?).
\item It would be interesting to study the D6 system as a BIon-like configuration on the world-volume of the D8, following the idea presented in \cite{Callan:1998} and briefly explored in the appendix.
\item As already remarked, one should really study the system for the non-antipodal D8-brane embedding, to be sure that the D6-brane is not dissolved on the D8.
Moreover, one could try and figure out how the system of D8 plus dissolved D6 looks like.
\item Eventually the main goal is to find a solution for the baryon, with disk-shaped sheet and the chiral edge mode turned on.
This configuration could be quantized to derive the baryon spectrum.
\item It would be interesting to work out a solution in the deconfined, chiral symmetry breaking phase \cite{Aharony:2006da}, using the black brane background.
\end{itemize}

%%%%%%%%%%%%%%%%%%%%%%%%%%%%%%%%
\vskip 15pt \centerline{\bf Acknowledgments} \vskip 10pt 

\noindent 
We are indebted with Ofer Aharony, Zohar Komargodski and Cobi Sonnenschein for their collaboration at the beginning of this project. We also thank Andrea Cappelli and Domenico Seminara for comments and very helpful discussions.

%%%%%%%%%%%%%%%%%%%%%%%%%%%%%%%%%%%%%%%%%%%%%%%%%%%
\appendix

\section{Including the scalar mode}
\label{appendixScalar}
\setcounter{equation}{0}
In section \ref{secSemiInfinite} we have ignored the scalar mode on the D8 world-volume parameterizing its transverse direction.
Here we consider its contribution at the lowest (quadratic) order (as we have done with the field strength). 
Let us call this mode $\phi$.
At leading order its effect is to shift the Lagrangian density as
\be
{\cal L}_0 \rightarrow {\cal L}_0 + \frac12 \sqrt{g} \partial_{M} \phi \partial^{M} \phi\,,
\ee
with $M,N = +,-,2,3,z$. We consider the ansatz
\be
 \phi=\phi(x_2,x_3,z,x_{\pm})\,.
\ee
The equation for the scalar mode
\begin{equation}
\partial_{M}\left(\sqrt{-g}\,g^{MN}\partial_{N}\phi\right) = 0,
\end{equation}
gives
\begin{equation*}
\partial_{z}\left(g(z)^{5/2}\partial_{z}\phi\right) + g(z)^{1/2}\left(\partial_{x_3}^{2} + \partial_{x_2}^{2}\right)\phi = 0.
\end{equation*}
We can try to solve this equation through a series expansion of $\phi$, as we did for the magnetic field strength. The ansatz for the solution reads
\begin{equation}
\phi = a(x_{\pm})\sum\limits_{n=0}^{\infty}\xi_{n}(z)\xi_{n}(0)J_{n}(r),
\end{equation}
with a generic function $a(x_{\pm})$ as found for the field strength of the charged mode. Then, the equation of motion becomes
\begin{align}
\sum\limits_{n=0}^{\infty}\left(\partial_{z}(g(z)^{5/2}\partial_{z}\xi_{n}(z))\xi_{n}(0)J_{n}(r) + g(z)^{1/2}\xi_{n}(z)\xi_{n}(0)\left(\partial_{x_3}^{2} + \partial_{x_2}^{2}\right)J_{n}(r)\right) = 0.
\end{align}
We can define $\widetilde{J}_{n}(r) \equiv \xi_{n}(0)J_{n}(r)$, in order to absorb the $n^{th}$ factor $\xi_{n}(0)$. The above equation can be solved requiring that $\xi_{n}(z)$ and $\widetilde{J}_{n}$ are solutions to the following equations
\begin{align}
\label{Gxi}
&\left(\left(\partial_{x_3}^{2} + \partial_{x_2}^{2}\right) - \tau_{n}\right)\widetilde{J}_{n}(r) = 0,\\\label{Gdexi}
&-g^{-1/2}(z)\partial_{z}(g(z)^{5/2}\partial_{z}\xi_{n}(z)) = \tau_{n}\xi_{n}(z).
\end{align}
The first one has a solution analogous to the $Y_{n}$:
\begin{equation}
\widetilde{J}_{n}(r) = \mathcal{N}K_{0}\left(\sqrt{\tau_{n}}r\right) \,\,\,\text{if}\,\,\, n > 0,\,\,\,\,\,\,\,\,\,\,\,\,\,\, \widetilde{J}_{0}(r) = -\mathcal{N}\ln(r)\,\,\, \text{if}\,\,\, n = 0,
\end{equation}
with $\mathcal{N}$ a normalization constant determined by requiring the on-shell action to be an effective action for the modes $\widetilde{J}_{n}$ canonically normalized. The eigenvalue equation for the functions $\xi_{n}(z)$ has to be addressed with the following orthonormality condition
\begin{align}
\int dz\, g(z)^{1/2}\xi_{n}(z)\xi_{m}(z) = \delta_{nm}.
\end{align}
The profile of the scalar grows big close to the line source, consistently with a BIon-like behavior. Of course, as for the gauge field, we cannot thrust the solution close to the source.

Then, we can compute the on-shell action for the scalar mode
\begin{equation}
S = -\kappa\sum\limits_{n=0}^{\infty}\int\,d^{4}x\,a(x_{\pm})\left( \partial_{i}\widetilde{J}_{n}(r)\partial_{i}\widetilde{J}_{n}(r) + \tau_{n}\widetilde{J}_{n}(r)\widetilde{J}_{n}(r)\right).
\end{equation}
This action looks just the same as \eqref{actionvortex},\footnote{The energy associated to this action has the same divergence as the tension of the vortex string.} but with different profiles $\widetilde{J}_{n}$. However, we have also the non trivial function $a(x_{\pm})$.  The BPS case, which is described in the paper \cite{Callan:1998} by Callan and Maldacena, is realized by taking $a(x_{\pm})$ constant. We leave this direction to
a future study.

%%%%%%%%%%%%%%%%%%%%%%%%%%%%%%%%%%%%%%%%%%%%%%%%%

\end{document}